\documentclass[a4paper,fleqn]{cas-sc}
\usepackage[numbers]{natbib}
\usepackage{booktabs}
\usepackage{adjustbox}
\usepackage{longtable}
\usepackage{caption}
\usepackage{mathtools}
\usepackage{amssymb}
\usepackage{amsmath}
\usepackage{etoolbox}

\newlength{\myalignwidth}

\def\tsc#1{\csdef{#1}{\textsc{\lowercase{#1}}\xspace}}
\tsc{WGM}
\tsc{QE}
\tsc{EP}
\tsc{PMS}
\tsc{BEC}
\tsc{DE}

\begin{document}
	\let\WriteBookmarks\relax
	\def\floatpagepagefraction{1}
	\def\textpagefraction{.001}
	\let\printorcid\relax
	
	\shorttitle{}
	
	\shortauthors{Chaoqian Wang {\it et~al.}}
	
	\title [mode = title]{Evolution of trust in structured populations}                      
	
	\author[1]{Chaoqian Wang}
	\cormark[1]
	\cortext[cor1]{Corresponding author}
	\ead{CqWang814921147@outlook.com}

	\address[1]{Department of Computational and Data Sciences, George Mason University, Fairfax, VA 22030, USA}

	\begin{abstract}
		The trust game, derived from an economics experiment, has recently attracted interest in the field of evolutionary dynamics. In a recent version of the evolutionary trust game, players adopt one of three strategies: investor, trustworthy trustee, or untrustworthy trustee. Trustworthy trustees enhance and share the investment with the investor, whereas untrustworthy trustees retain the full amount, betraying the investor. Following this setup, we investigate a two-player trust game, which is analytically feasible under weak selection. We explore the evolution of trust in structured populations, factoring in four strategy updating rules: pairwise comparison (PC), birth-death (BD), imitation (IM), and death-birth (DB). Comparing structured populations with well-mixed populations, we arrive at two main conclusions. First, in the absence of untrustworthy trustees, there is a saddle point between investors and trustworthy trustees, with collaboration thriving best in well-mixed populations. The collaboration diminishes sequentially from DB to IM to PC/BD updating rules in structured populations. Second, an invasion of untrustworthy trustees makes this saddle point unstable and leads to the extinction of investors. The 3-strategy system stabilizes at an equilibrium line where the trustworthy and untrustworthy trustees coexist. The stability span of trustworthy trustees is maximally extended under the PC and BD updating rules in structured populations, while it decreases in a sequence from IM to DB updating rules, with the well-mixed population being the least favorable. This research thus adds an analytical lens to the evolution of trust in structured populations.
	\end{abstract}
	
	
	

	\maketitle
	
	\section{Introduction}\label{sec_intro}
	The ``trust game'' is an important experiment in behavioral economics utilized to investigate trust behavior between individuals. Developed initially in the 1990s, it involves two anonymous players, known as the ``trustor'' and the ``trustee''~\cite{berg1995trust}. The trustor is endowed with a monetary sum and faces the decision to send a part or the entirety of this sum to the trustee. The transferred amount multiplies, augmenting the value before it reaches the trustee. The trustee then decides the portion of the increased sum to retain and what fraction to return to the trustor. The theoretical optimal strategy from a pure economic standpoint is for the trustor not to transfer any amount, anticipating that the trustee aims to maximize their payoff and therefore will not return any amount. However, the game often leads to the result of trust and reciprocity, where the trustee returns a part of the money. 
	
	Traditionally, the trust game has been explored in various economic experimental setups, examining strategies with incomplete information~\cite{anderhub2002experimental} and over repeated interactions~\cite{engle2004evolution}. The richness of behaviors observed has been explored further in research where individuals embrace both roles of the trustor and trustee~\cite{burks2003playing}. With meta-data analysis, it has offered a broadened perspective on strategy formulation~\cite{johnson2011trust}. The landscape of trust games extends beyond economic rationales, touching upon broader humanity factors such as gender and culture~\cite{croson1999gender}, alongside the subjective territories of beauty and expectations~\cite{wilson2006judging}. Moreover, biological investigations have unraveled genetic influences on trust dynamics~\cite{cesarini2008heritability}, albeit these represent but a fraction of the multidimensional trust game investigations. The reader can refer to a recent review~\cite{alos2019trust} that provides a more complete picture of the development of the traditional trust games over the last two decades.
	
	Evolutionary dynamics offers a fresh angle on traditional game and behavior theory, including honesty~\cite{capraro2019evolution,capraro2020lying,kumar2021evolution}, morality~\cite{capraro2018grand,capraro2021mathematical}, and trust~\cite{abbass2015n,chica2017networked}, among others. As a move beyond the traditional theory, evolutionary dynamics illustrates how players, while seeking higher individual payoffs, can still choose to cooperate, favoring the collective's interest over their own highest payoff~\cite{nowak2006evolutionary,sigmund2010calculus}. This cooperative behavior is amplified in structured populations that closely mirror real-world scenarios where individuals interact with their neighbors rather than the entire population, fostering ``spatial reciprocity'' through local strategy interaction and reproduction~\cite{nowak1992evolutionary,szabo1998evolutionary}. Evolutionary graph theory analyzes this phenomenon, studying the dynamics of evolution in such structured populations~\cite{lieberman2005evolutionary,szabo2007evolutionary,nowak2010evolutionary}. A commonly used theory is the pair approximation approach, which introduces the marginal effect of games based on the Voter model~\cite{clifford1973model}. Originating from general biology and physics fields~\cite{gutowitz1987local,matsuda1987lattice,szabo1991correlations,matsuda1992statistical}, early pair approximation methodologies found application in evolutionary games on regular graphs, revealing the `$b/c>k$' rule~\cite{ohtsuki2006simple}. Soon after, the methodology was applied for multi-strategy two-player games, spawning the concept of ``replicator dynamics on graphs''~\cite{ohtsuki2006replicator}, which supplements the traditional replicator dynamics in well-mixed populations~\cite{taylor1978evolutionary}. Recent advancements in the pair approximation approach have facilitated studies on game transitions~\cite{su2019evolutionary} and asymmetric social interactions in evolutionary dynamics~\cite{su2022evolutionasym}. A parallel branch in evolutionary graph theory is the identity-by-descent theory, capable of depicting evolutionary dynamics across any network structure~\cite{allen2014games,allen2017evolutionary}. This framework has spurred significant findings recently, including~\cite{su2019spatial,mcavoy2020social,su2022evolution,su2023strategy,wang2023evolution,wang2023inertia,wang2023conflict,wang2023greediness}. For a comprehensive understanding of the progress in evolutionary dynamics over the past two decades, the reader can refer to a recent review~\cite{perc2017statistical}.
	
	Other than a few earlier works~\cite{masuda2012coevolution,tarnita2015fairness}, one seminal work applying evolutionary dynamics to trust games began with the original $N$-person trust game in a well-mixed and infinite population~\cite{abbass2015n} (and its lesser-known version in a finite population~\cite{greenwood2016finite}). In the $N$-person trust game, players are allowed to employ three strategies: investor (trustor), trustworthy trustee, and untrustworthy trustee. For simplicity, the option of not investing is unavailable to investors. The payoffs are calculated based on the traditional trust game and are proceeded through the lens of evolutionary dynamics. In recent years, the underlying model of the $N$-person trust game has inspired a wide range of studies, including consideration of network structures~\cite{chica2017networked}, punishment~\&~reward mechanisms~\cite{chica2019evolutionary,fang2021evolutionary,sun2022evolution}, reputations~\cite{hu2021adaptive,li2022n,xia2022costly}, diverse investment patterns~\cite{shang2023evolutionary}, conditional investment in repeated interactions~\cite{liu2022conditional}, comparison with logit dynamics~\cite{towers2017n}, and the effects of different updating rules~\cite{chica2019effects}. Some studies have adopted alternative strategy settings, moving away from the 3-strategy approach. Examples include a simulation study on a square lattice~\cite{kumar2020evolution}, research on fixed provider and consumer roles~\cite{chiong2022evolution,liu2023n,guo2023evolution}, and further theoretical studies~\cite{lim2020stochastic,lim2021synergy,lim2023trust}.
	
	There have been studies in the literature on the evolutionary graph theory of trust games. In particular, Lim and Capraro~\cite{lim2021synergy} have theoretically studied a 4-strategy system in structured populations, where a player can be an investor and trustee simultaneously. However, the seminal 3-strategy system has yet to apply the theoretical approach of evolutionary graph theory. Specifically, previous studies have either been based on analytical studies of well-mixed populations~\cite{abbass2015n,fang2021evolutionary} or Monte Carlo simulation studies in structured populations~\cite{chica2017networked,hu2021adaptive,li2022n,xia2022costly}. Analytical studies on structured populations are still lacking. Ohtsuki and Nowak~\cite{ohtsuki2006replicator} have provided the general replicator equations for multi-strategy two-player games on regular graphs, where each player has the same number of neighbors. This enables the theoretical analysis of trust games in a structured population. Given that most current studies on evolutionary trust games are $N$-player games, which are not analytically feasible at the moment, we need to first transfer the underlying model to the form of two-player games. In this work, we follow the 3-strategy setup of the seminal work on evolutionary trust games~\cite{abbass2015n}, but propose a corresponding two-player version. On this basis, we focus on theoretical solutions under different strategy updating rules in structured populations and how they affect the evolution of trust in the structured population. We start by introducing a corresponding two-player trust game model in the next section.
	
	\section{Model}
	\subsection{Trust game}\label{sec_trust}
	The two-player trust game that we propose employs a 3-strategy system. A player can employ one of the following strategies:
	
	1. Investor ($I$), also known as a trustor, who invests $t_V$ to the trustee co-player and expects a return from the trustee, which is conditional upon the trustee being trustworthy or not. Here, $t_V>0$ is the trusted value, an input parameter. If the co-player also adopts the investor role, neither invests, resulting in no action.
	
	2. Trustworthy trustee ($T$), who multiplies the investment from the investor by $R_T$ ($R_T>1$). The trustee receives $R_T t_V$. The investor also receives $R_T t_V$. If the co-player is also a trustee, no transaction occurs.
	
	3. Untrustworthy trustee ($U$), who multiplies the investment from the investor by $R_U$ ($R_U>R_T$). The trustee receives $R_U t_V$. The investor receives nothing. If the co-player is also a trustee (either $T$ or $U$), no transaction takes place.
	
	Based on the described strategies, we construct the following payoff matrix:
	\begin{equation}\label{eq_aij}
		\left[a_{ij}\right]=
		\begin{pmatrix}
			0 & -t_V+R_T t_V & -t_V \\[0.5em]
			R_T t_V & 0 & 0 \\[0.5em]
			R_U t_V & 0 & 0
		\end{pmatrix}.
	\end{equation}
	An $I$-player gains $-t_V+R_T t_V$ when interacting with a $T$-player and incurs loss of $-t_V$ when facing a $U$-player. On the other hand, a $T$-player earns $R_T t_V$ in encounters with an $I$-player, while a $U$-player secures $R_U t_V$ against an $I$-player.
	
	The parameters $R_T$ and $R_U$ yield specific relations. We have mentioned $R_T>1$, which guarantees that an investment to a trustworthy trustee always brings a positive return ($-t_V+R_T t_V>0$). Also, $R_U>R_T$ indicates that being untrustworthy can yield greater benefits compared to being trustworthy ($R_U t_V>R_T t_V$). Moreover, we must require $R_U<2R_T$, which means being trustworthy is always a prosocial behavior that confers greater collective benefits ($-t_V+R_T t_V+R_T t_V>-t_V+R_U t_V$). To sum up, the relation $1<R_T<R_U<2R_T$ is established.
	
	We also notice that $t_V$ can be extracted from the payoff matrix $\left[a_{ij}\right]$, which is similar to the cost parameter $c$ in public goods games~\cite{wang2023inertia,wang2023zealous}. In other words, $t_V$ serves a role analogous to that of the selection strength (introduced in Section~\ref{sec_evolution}). The effect of $t_V$ is only visible when selection strength is non-marginal, which does not apply to this study. Therefore, we can simply set $t_V=1$ and do not explore it.
	
	\subsection{Evolutionary dynamics}\label{sec_evolution}
	During each elementary step, a random focal player is selected to update its strategy through the interactions with its $k$ neighbors. If the population is well-mixed, these neighbors are randomly chosen from the population and vary over time. If the population is structured, the neighbors remain constant. All players have the same number of neighbors, denoted as $k$.
	
	The randomly selected focal player earns a payoff from playing $k$ trust games given in Section~\ref{sec_trust} against $k$ neighbors. The neighbors similarly determine their own payoffs through interactions with their respective neighbors. The focal player adopts the strategy ($I$, $T$, or $U$) of a neighbor or retains its own strategy, based on which strategy yields the higher payoff. The more successful strategy is more likely to be adopted, through a process further detailed under various updating rules in Section~\ref{sec_struc}. We assume a weak selection strength, indicating that the differences in payoff have only a marginal influence on the evolutionary dynamics.
	
	The weak selection framework allows for the analysis of dynamics in structured populations. In light of this, we analyze the evolutionary dynamics discussed earlier using replicator dynamics in both well-mixed~\cite{taylor1978evolutionary} and structured populations~\cite{ohtsuki2006replicator}. Although the focus of this work is on structured populations, we begin our analysis with well-mixed populations to establish a basis for comparison.
	
	\section{The well-mixed population}\label{sec_wellmix}
	In an infinite well-mixed population, the frequency of $i$-players is denoted by $x_i$, where $\sum_i x_i=1$, $i=I,T,U$. The system state can be described by $\mathbf{x}=(x_I,x_T,x_U)$. The replicator equations are $\dot{x}_i=x_i(f_i-\phi)$, where $f_i=\sum_{j}x_j a_{ij}$ is the mean payoff of $i$-players, $\phi=\sum_i x_i f_i$ is the mean payoff of the population~\cite{taylor1978evolutionary}. From Eq.~(\ref{eq_aij}), we obtain
	\begin{subequations}\label{eq_f}
		\begin{alignat}{2}
			&f_I&&=[x_T(R_T-1)-x_U]t_V, \\
			&f_T&&=x_I R_T t_V, \\
			&f_U&&=x_I R_U t_V,
		\end{alignat}
	\end{subequations}
	and
	\begin{equation}\label{eq_phi}
		\phi=x_I[x_T(R_T-1)-x_U]t_V+x_I t_V(x_T R_T+x_U R_U).
	\end{equation}
	Therefore, the replicator dynamics for the well-mixed population is
	\begin{equation}\label{eq_replicator_wm}
		\left\{
		\begin{array}{@{\hspace{0.1em}}l@{\hspace{0.25em}}l}
			\dot{x}_I &= x_I t_V [x_T (R_T-1)-x_U-x_I x_T (2R_T-1)-x_I x_U (R_U-1)], \\[0.5em]
			\dot{x}_T &= x_I x_T t_V [R_T-x_T (2R_T-1)-x_U (R_U-1)], \\[0.5em]
			\dot{x}_U &= x_I x_U t_V [R_U-x_T (2R_T-1)-x_U (R_U-1)].
		\end{array}
		\right.
	\end{equation}
	We can see that the replicator equations in well-mixed populations are independent of $k$.
	
	Solving for $\dot{\mathbf{x}}=\mathbf{0}$ yields two distinct equilibrium points and one equilibrium line in the system described by Eq.~(\ref{eq_replicator_wm}). The first equilibrium point is the $I$-vertex, $\mathbf{x}^{(I)}=(1,0,0)$. The second equilibrium point is located on the $IT$-edge, 
	\begin{equation}\label{eq_xit_wm}
		\mathbf{x}^{(IT)}=\left(
		\frac{R_T-1}{2R_T-1}, \frac{R_T}{2R_T-1}, 0
		\right).
	\end{equation}
	The equilibrium line encompasses the entire $TU$-edge,
	\begin{equation}
		\mathbf{x}^{(TU)}=\left(
		0, x_T^{(TU)}, x_U^{(TU)}
		\right),
	\end{equation}
	where $0\leq x_T^{(TU)}, x_U^{(TU)}\leq 1$, $x_T^{(TU)}+x_U^{(TU)}=1$. On the line represented by $\mathbf{x}^{(TU)}$, there are infinite equilibrium points, including the $T$- and $U$-vertices, $(0,1,0)$ and $(0,0,1)$.
	
	According to the stability analysis (see Appendix~\ref{sec_appen}), the equilibrium point $\mathbf{x}^{(I)}$ is unstable. The equilibrium point $\mathbf{x}^{(IT)}$ is a saddle point, being only stable along the $IT$-edge and turning unstable when any $U$-player is introduced into the system. The equilibrium line $\mathbf{x}^{(TU)}$ remains stable only within a certain interval. More precisely, the equilibrium line $\mathbf{x}^{(TU)}$ remains stable when $x_{T}^{(TU)}<x_{T,\star}^{(TU)}$, where
	\begin{equation}\label{eq_xtu_wm}
		x_{T,\star}^{(TU)}=\frac{1}{R_T}.
	\end{equation}
	
\begin{figure}\centering
		\includegraphics[width=\linewidth]{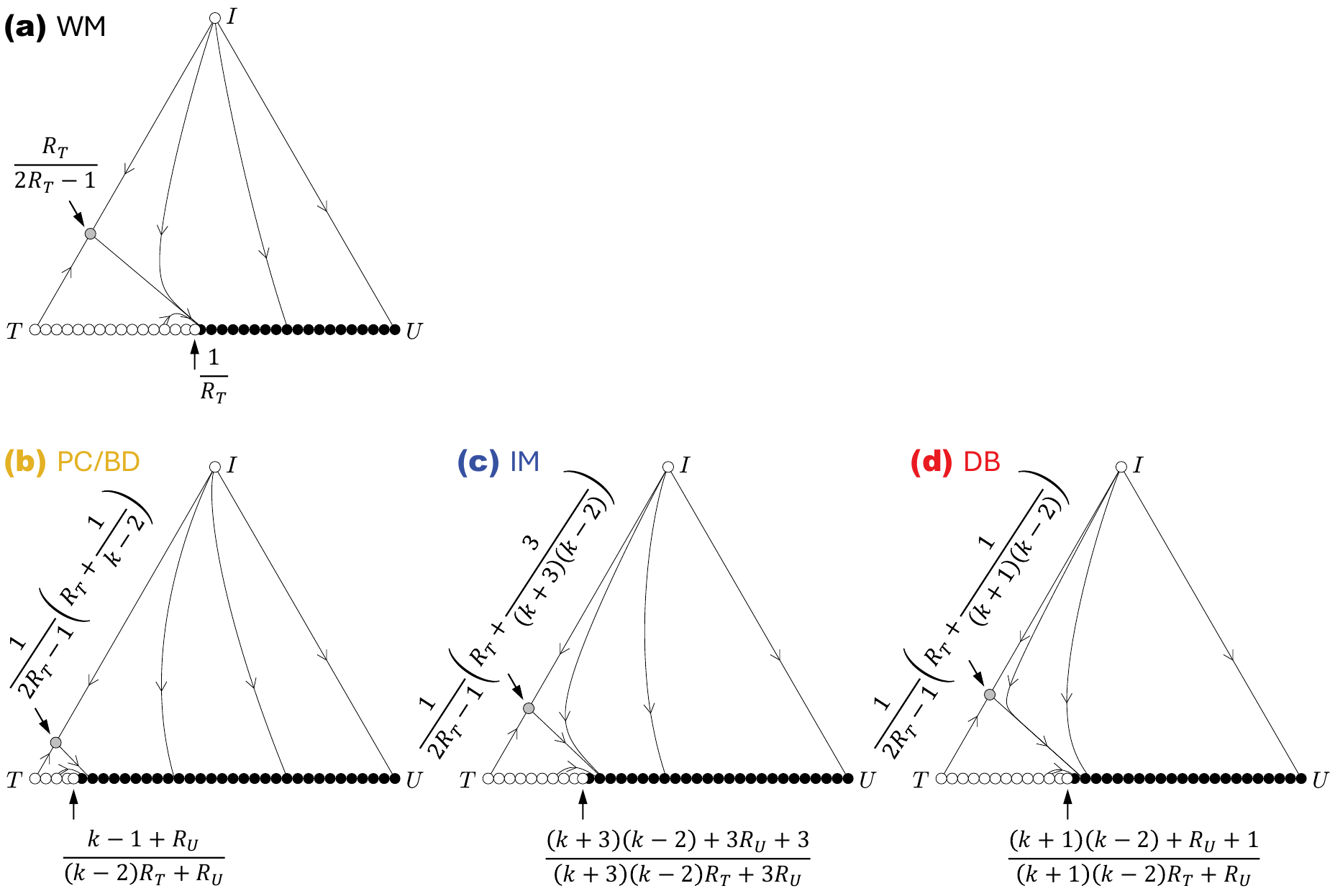}
		\caption{The evolution of the system depicted by ternary diagrams, including (\textbf{a}) well-mixed (WM) and structured populations under the (\textbf{b}) PC/BD, (\textbf{c}) IM, and (\textbf{d}) DB updating rules. The black points are stable, the white points are unstable, and the grey point indicates a saddle point. On the $IT$-edge, the grey point marks $\mathbf{x}^{(IT)}$, and the annotated formula is the analytical expression of $x_T^{(IT)}$. On the $TU$-edge, a particular critical white point marks the boundary separating stable and unstable regions along the equilibrium line, and the annotated formula is the analytical expression of $x_{T,\star}^{(TU)}$. Input parameters: $R_T=1.8$, $R_U=2$, $t_V=1$, $k=4$.}\label{fig1}
	\end{figure}

	Refer to Fig.~\ref{fig1}(a) for a numerical demonstration of these equilibrium points and their respective stability.
	
	\section{The structured population}\label{sec_struc}
	According to~\cite{ohtsuki2006replicator}, the essence of evolutionary dynamics in structured populations entails a transformation of the payoff matrix $\left[a_{ij}\right]\gets \left[a_{ij}+b_{ij}\right]$ in comparison to well-mixed populations. More precisely, the replicator dynamics becomes $\dot{x}_i=x_i(f_i+g_i-\phi)$, where $g_i=\sum_{j} x_j b_{ij}$ is the additional advantage for $i$-players brought by the network structure. Here, $b_{ij}$ depends on specific strategy updating rules. The rules under consideration include pairwise comparison (PC), birth-death (BD), imitation (IM), and death-birth (DB). Below, we discuss them separately.
	
	\subsection{Pairwise comparison (PC) and birth-death (BD)}
	The replicator dynamics of the PC and BD updating rules in a structured population are equivalent to each other under weak selection~\cite{ohtsuki2006replicator}.
	
	In the PC updating rule, a focal player and one of its neighbors are randomly selected. With a probability marginally proportional to the payoff in the pair, the focal player adopts the strategy of the selected neighbor or keeps its own strategy~\cite{szabo1998evolutionary}. In the BD updating rule, a focal player is selected with a probability marginally proportional to the payoff among the population, then a random neighbor adopts the focal player's strategy~\cite{lieberman2005evolutionary}.
	
	According to~\cite{ohtsuki2006replicator}, both rules adhere to the following formula for calculating $b_{ij}$:
	\begin{equation}\label{eq_bPC}
		b_{ij}=\frac{a_{ii}+a_{ij}-a_{ji}-a_{jj}}{k-2}.
	\end{equation}
	Using Eq.~(\ref{eq_aij}), we express each element in the matrix $\left[b_{ij}\right]$ as
	\begin{equation}
		\left[b_{ij}\right]=\frac{1}{k-2}
		\begin{pmatrix}
			0 & -t_V & -t_V-R_U t_V \\[0.5em]
			t_V & 0 & 0 \\[0.5em]
			R_U t_V+t_V & 0 & 0
		\end{pmatrix}.
	\end{equation}
	Therefore, $g_i=\sum_{j} x_j b_{ij}$ is computed as
	\begin{subequations}
		\begin{alignat}{2}
			&g_I &&= \displaystyle{-\frac{x_T+x_U(R_U-1)}{k-2}t_V}, \\
			&g_T &&= \displaystyle{\frac{x_I}{k-2}t_V}, \\
			&g_U &&= \displaystyle{\frac{x_I(R_U+1)}{k-2}t_V}.
		\end{alignat}
	\end{subequations}
	
	The resulting replicator dynamics in structured populations $\dot{x}_i=x_i(f_i+g_i-\phi)$ under the pairwise comparison or birth-death updating rule is as follows:
	\begin{equation}\label{eq_replicator_pc}
		\left\{
		\begin{array}{@{\hspace{0.1em}}l@{\hspace{0.25em}}l}
			\dot{x}_I &= \displaystyle{x_I t_V \left[x_T (R_T-1)-x_U-\frac{x_T+x_U(R_U+1)}{k-2}-x_I x_T (2R_T-1)-x_I x_U (R_U-1)\right]}, \\[1em]
			\dot{x}_T &= \displaystyle{x_I x_T t_V \left[R_T+\frac{1}{k-2}-x_T (2R_T-1)-x_U (R_U-1)\right]}, \\[1em]
			\dot{x}_U &= \displaystyle{x_I x_U t_V \left[R_U+\frac{R_U+1}{k-2}-x_T (2R_T-1)-x_U (R_U-1)\right]}.
		\end{array}
		\right.
	\end{equation}
	
	\subsection{Imitation (IM)}
	In the IM updating rule, a focal player is randomly selected. With a probability marginally proportional to the payoff among all neighbors and itself, the focal player adopts the strategy of a neighbor or keeps its own strategy~\cite{nowak1992evolutionary}.
	
	According to~\cite{ohtsuki2006replicator}, the IM rule adheres to the following formula for calculating $b_{ij}$:
	\begin{equation}\label{eq_bIM}
		b_{ij}=\frac{(k+3)a_{ii}+3a_{ij}-3a_{ji}-(k+3)a_{jj}}{(k+3)(k-2)}.
	\end{equation}
	From this formula, we can calculate
	\begin{equation}
		\left[b_{ij}\right]=\frac{3}{(k+3)(k-2)}
		\begin{pmatrix}
			0 & -t_V & -t_V-R_U t_V \\[0.5em]
			t_V & 0 & 0 \\[0.5em]
			R_U t_V+t_V & 0 & 0
		\end{pmatrix},
	\end{equation}
	and
	\begin{subequations}
		\begin{alignat}{2}
			&g_I &&= \displaystyle{-\frac{3x_T+3x_U(R_U-1)}{(k+3)(k-2)}t_V}, \\
			&g_T &&= \displaystyle{\frac{3x_I}{(k+3)(k-2)}t_V}, \\
			&g_U &&= \displaystyle{\frac{3x_I(R_U+1)}{(k+3)(k-2)}t_V}.
		\end{alignat}
	\end{subequations}
	
	Therefore, the replicator dynamics in structured populations under the imitation updating rule is
	\begin{equation}\label{eq_replicator_im}
		\left\{
		\begin{array}{@{\hspace{0.1em}}l@{\hspace{0.25em}}l}
			\dot{x}_I &= \displaystyle{x_I t_V \left[x_T (R_T-1)-x_U-\frac{3x_T+3x_U(R_U+1)}{(k+3)(k-2)}-x_I x_T (2R_T-1)-x_I x_U (R_U-1)\right]}, \\[1em]
			\dot{x}_T &= \displaystyle{x_I x_T t_V \left[R_T+\frac{3}{(k+3)(k-2)}-x_T (2R_T-1)-x_U (R_U-1)\right]}, \\[1em]
			\dot{x}_U &= \displaystyle{x_I x_U t_V \left[R_U+\frac{3(R_U+1)}{(k+3)(k-2)}-x_T (2R_T-1)-x_U (R_U-1)\right]}.
		\end{array}
		\right.
	\end{equation}
	
	\subsection{Death-birth (DB)}
	In the DB updating rule, a focal player is randomly selected. With a probability marginally proportional to the payoff among all neighbors, the focal player adopts the strategy of a neighbor~\cite{ohtsuki2006simple}. Compared to the IM rule, the DB rule completely ignores the payoff of the focal player, making it unable to retain its own strategy~\cite{wang2023evolution,wang2023inertia}.
	
	According to~\cite{ohtsuki2006replicator}, the DB rule adheres to the following formula for calculating $b_{ij}$:
	\begin{equation}\label{eq_bDB}
		b_{ij}=\frac{(k+1)a_{ii}+a_{ij}-a_{ji}-(k+1)a_{jj}}{(k+1)(k-2)},
	\end{equation}
	by which we compute
	\begin{equation}
		\left[b_{ij}\right]=\frac{1}{(k+1)(k-2)}
		\begin{pmatrix}
			0 & -t_V & -t_V-R_U t_V \\[0.5em]
			t_V & 0 & 0 \\[0.5em]
			R_U t_V+t_V & 0 & 0
		\end{pmatrix},
	\end{equation}
	and
	\begin{subequations}
		\begin{alignat}{2}
			&g_I &&= \displaystyle{-\frac{x_T+x_U(R_U-1)}{(k+1)(k-2)}t_V}, \\
			&g_T &&= \displaystyle{\frac{x_I}{(k+1)(k-2)}t_V}, \\
			&g_U &&= \displaystyle{\frac{x_I(R_U+1)}{(k+1)(k-2)}t_V}.
		\end{alignat}
	\end{subequations}
	
	In this way, the replicator dynamics in structured populations under the death-birth updating rule is
	\begin{equation}\label{eq_replicator_db}
		\left\{
		\begin{array}{@{\hspace{0.1em}}l@{\hspace{0.25em}}l}
			\dot{x}_I &= \displaystyle{x_I t_V \left[x_T (R_T-1)-x_U-\frac{x_T+x_U(R_U+1)}{(k+1)(k-2)}-x_I x_T (2R_T-1)-x_I x_U (R_U-1)\right]}, \\[1em]
			\dot{x}_T &= \displaystyle{x_I x_T t_V \left[R_T+\frac{1}{(k+1)(k-2)}-x_T (2R_T-1)-x_U (R_U-1)\right]}, \\[1em]
			\dot{x}_U &= \displaystyle{x_I x_U t_V \left[R_U+\frac{R_U+1}{(k+1)(k-2)}-x_T (2R_T-1)-x_U (R_U-1)\right]}.
		\end{array}
		\right.
	\end{equation}

	\subsection{Results}
	By solving $\dot{\mathbf{x}}=\mathbf{0}$ in Eqs.~(\ref{eq_replicator_pc}), (\ref{eq_replicator_im}), and (\ref{eq_replicator_db}), we obtain the equilibrium points in the structured population under the PC/BD, IM, and DB rules. They share similar stability properties, with minor differences in the analytic forms. We identify two distinct equilibrium points and one equilibrium line. The first equilibrium point is the $I$-vertex, $\mathbf{x}^{(I)}=(1,0,0)$. The second equilibrium point is on the $IT$-edge, 
	\begin{equation}\label{eq_xit_st}
		\mathbf{x}^{(IT)}=\frac{1}{2R_T-1}
		\begin{cases}
			\displaystyle{\left(
				R_T-1-\frac{1}{k-2}, 
				R_T+\frac{1}{k-2}, 0
				\right)}, &\mbox{PC/BD,} \\[1em]
			\displaystyle{\left(
				R_T-1-\frac{3}{(k+3)(k-2)}, 
				R_T+\frac{3}{(k+3)(k-2)}, 0
				\right)}, &\mbox{IM,} \\[1em]
			\displaystyle{\left(
				R_T-1-\frac{1}{(k+1)(k-2)}, 
				R_T+\frac{1}{(k+1)(k-2)}, 0
				\right)}, &\mbox{DB.}
		\end{cases}
	\end{equation}
	The equilibrium line represents the whole $TU$-edge,
	\begin{equation}
		\mathbf{x}^{(TU)}=\left(
		0, x_T^{(TU)}, x_U^{(TU)}
		\right),
	\end{equation}
	where $0\leq x_T^{(TU)}, x_U^{(TU)}\leq 1$, $x_T^{(TU)}+x_U^{(TU)}=1$. There exist infinite equilibrium points on the line depicted by $\mathbf{x}^{(TU)}$, including the $T$-vertex at $(0,1,0)$ and the $U$-vertex at $(0,0,1)$.
	
	As the stability analysis indicates (see Appendix~\ref{sec_appen}), the equilibrium point $\mathbf{x}^{(I)}$ is unstable. The equilibrium point $\mathbf{x}^{(IT)}$ is a saddle point, which is stable only along the $IT$-edge and becomes unstable when any U-player is introduced into the system. The equilibrium line $\mathbf{x}^{(TU)}$ is stable only within the interval of $x_{T}^{(TU)}<x_{T,\star}^{(TU)}$, where
	\begin{equation}\label{eq_xtu_st}
		x_{T,\star}^{(TU)}=\begin{cases}
			\displaystyle{\frac{k-1+R_U}{(k-2)R_T+R_U}}, &\mbox{PC/BD,} \\[1em]
			\displaystyle{\frac{(k+3)(k-2)+3R_U+3}{(k+3)(k-2)R_T+3R_U}}, &\mbox{IM,} \\[1em]
			\displaystyle{\frac{(k+1)(k-2)+R_U+1}{(k+1)(k-2)R_T+R_U}}, &\mbox{DB.}
		\end{cases}
	\end{equation}
	
	See Fig.~\ref{fig1}(b), (c), and (d) for a numerical demonstration of these equilibrium points and their stability under the PC/BD, IM, and DB rules, respectively.
	
	\section{Discussion}
	In Sections~\ref{sec_wellmix} and \ref{sec_struc}, we obtained the dynamics in both the well-mixed and structured populations. These scenarios can be compared from two perspectives: the characteristics of the saddle point $\mathbf{x}_T^{(IT)}$ on the $IT$-edge, and the stability interval of the equilibrium line $\mathbf{x}_T^{(TU)}$ on the $TU$-edge. In the subsequent sections, we will discuss them separately.
	
	\subsection{The $IT$-edge}
	In the absence of $U$-players, the system behavior on the $IT$-edge describes the competition and collaboration between the investors ($I$) and the trustworthy trustees ($T$). From the payoff matrix, these two strategies must form an interdependent relationship to generate a payoff; neither the investor nor the trustee can do so on their own, necessitating collaboration to realize positive outcomes. This is similar to the interdependent mechanism between different cooperative strategies~\cite{wang2021public}. We can deduce the optimal system state where the payoff of the population is maximized. Let us set $x_U=0$ and $x_I=1-x_T$ in Eq.~(\ref{eq_phi}), which thus becoming a function of $x_T$,
	\begin{equation}\label{eq_phixt}
		\phi=(1-x_T) f_I+x_T f_T=x_T(1-x_T)(2R_T-1).
	\end{equation}
	According to Eq.~(\ref{eq_phixt}), the mean payoff of the population $\phi$ is maximized at $x_T=1/2$. That is, when investors and trustworthy trustees are each half of the population, the system state best serves the collective interest. Thus, we can define the equilibrium point close to $x_T^{(IT)}=1/2$ as the point of prosocial behavior, characterized by collaboration between investors and trustworthy trustees.
	
	Based on this criterion, we compare the $x_T^{(IT)}$ values given by Eqs.~(\ref{eq_xit_wm}) and (\ref{eq_xit_st}) in the four different systems, and find that
	\begin{equation}
		\frac{1}{2R_T-1}\left(R_T+\frac{1}{k-2}\right)>
		\frac{1}{2R_T-1}\left(R_T+\frac{3}{(k+3)(k-2)}\right)
		>\frac{1}{2R_T-1}\left(R_T+\frac{1}{(k+1)(k-2)}\right)>
		\frac{R_T}{2R_T-1}>
		\frac{1}{2}.
	\end{equation}
	That is, PC/BD>IM>DB>WM, and WM>1/2. The pairwise comparison and birth-death updating rules lead to the largest $x_T^{(IT)}$ value, and the imitation, death-birth, and the well-mixed population lead to smaller $x_T^{(IT)}$ values sequentially. In the smallest case, the well-mixed population, we still have $x_T^{(IT)}>1/2$, and $x_T^{(IT)}\to 1/2$ only as $R_T\to \infty$. This naturally follows from the fact that a trustee always gains a higher payoff than an investor, $R_T t_V>-t_V+R_T t_V$. Therefore, even if the simultaneous presence of both is necessary to generate payoff for both strategies, being a trustee is a more attractive option.
	
	From this perspective, it is evident that the well-mixed population is the most favorable for maintaining the trustee ratio near $1/2$, implying that it is most favorable for investor-trustee collaboration to thrive. On the contrary, the structured population is not conducive to maintaining the trustee ratio near $1/2$, with the PC and BD updating rules being the most harmful to collaboration.
	
	The essence of spatial reciprocity brought by network structures is that players with the same strategy have a higher chance of meeting with each other. In the calculation of the additional advantage $b_{ij}$ brought by the network structure in Eqs.~(\ref{eq_bPC}), (\ref{eq_bIM}), and (\ref{eq_bDB}), we notice that $a_{ii}-a_{jj}=0$ always holds for the different updating rules. There is no advantage when an investor or a trustee meet with the same strategy type. Instead, as we have discussed, they must collaborate with the opposing strategy type to generate a payoff. This is the reason why structured populations disfavor collaboration between the investor and trustworthy trustees compared to a well-mixed population.
	
\begin{figure}\centering
		\includegraphics[width=\linewidth]{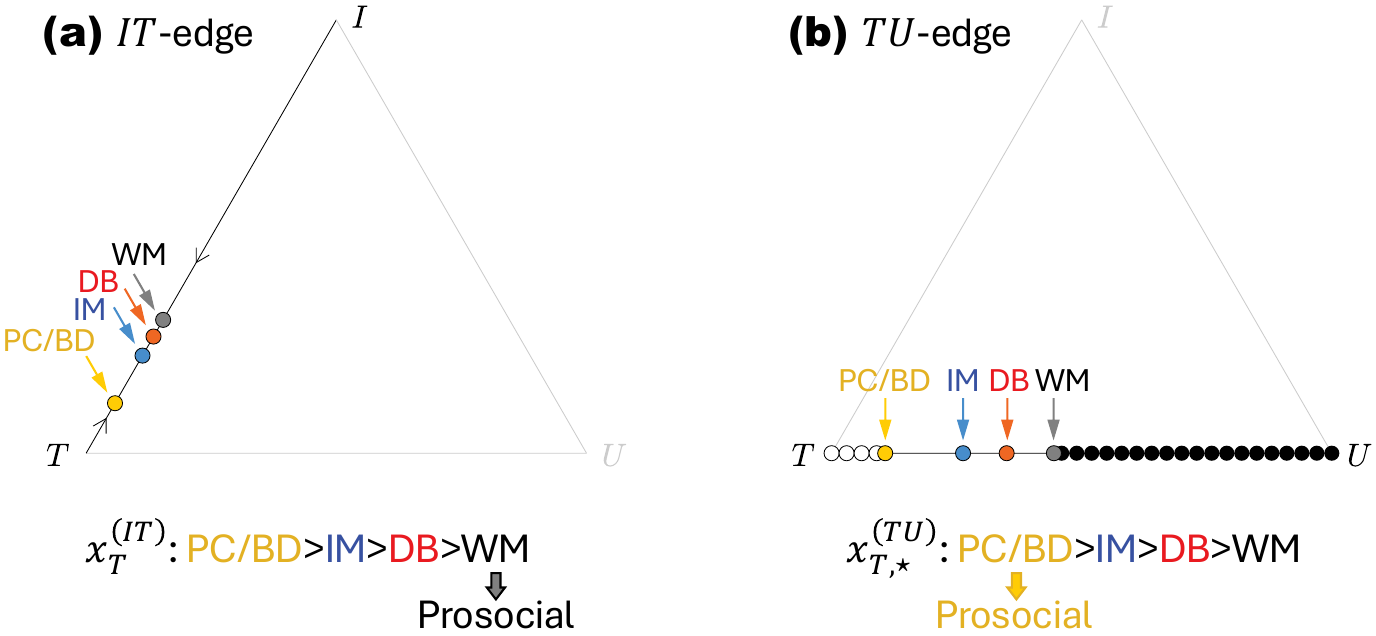}
		\caption{Comparison between the well-mixed (WM) and structured populations under the PC/BD, IM, and DB updating rules. (\textbf{a}) Comparison of the saddle point along the $IT$-edge. Since $x_T^{(IT)}=1/2$ is most beneficial to the population, it can be concluded that the well-mixed population can favor the prosocial behavior of collaboration more than structured populations. (\textbf{b}) Comparison of the critical point along the $TU$-edge. Since trustworthy trustees are prosocial and the interval where $x_{T}^{(TU)}<x_{T,\star}^{(TU)}$ is stable, the PC/BD updating in structured populations can enlarge the attraction interval and favor the prosocial behavior of trust. Input parameters: $R_T=1.8$, $R_U=2$, $t_V=1$, $k=4$.}\label{fig2}
	\end{figure}

	Figure~\ref{fig2}(a) numerically compares the equilibrium points of the well-mixed population and the structured population with different updating rules on the $IT$-edge, which visually demonstrates the observations discussed above. 
	
	\subsection{The $TU$-edge}
	While the investors and trustworthy trustees form prosocial collaboration and reciprocity, an invasion of untrustworthy trustees can break this state and lead to the extinction of investors; that is, the saddle point $\mathbf{x}_T^{(IT)}$ on the $IT$-edge becomes unstable in the direction of $U$-vertex. As a result, the evolution ends with the coexistence of trustworthy and untrustworthy trustees, with no investors, stabilizing at the equilibrium line on the $TU$-edge. In this case, we define a state with more trustworthy trustees as a better ending. Since the equilibrium line is stable at the interval $x_{T}^{(TU)}<x_{T,\star}^{(TU)}$, a larger critical value $x_{T,\star}^{(TU)}$ indicates an extended stable interval with more trustworthy trustees.
	
	Based on this criterion, we compare the $x_{T,\star}^{(TU)}$ values given by Eqs.~(\ref{eq_xtu_wm}) and (\ref{eq_xtu_st}) in the four different systems, and find that
	\begin{equation}\label{eq_tudiscuss}
		\frac{k-1+R_U}{(k-2)R_T+R_U}>
		\frac{(k+3)(k-2)+3R_U+3}{(k+3)(k-2)R_T+3R_U}>
		\frac{(k+1)(k-2)+R_U+1}{(k+1)(k-2)R_T+R_U}>
		\frac{1}{R_T}.
	\end{equation}
	That is, PC/BD>IM>DB>WM. The pairwise comparison and birth-death updating rules lead to the largest $x_{T,\star}^{(TU)}$ value and favor trustworthy trustees the most. The imitation, death-birth, and the well-mixed population lead to smaller $x_{T,\star}^{(TU)}$ values sequentially. In this sense, the PC and BD updating rules in structured populations are most favorable to support trustworthy trustees against untrustworthy trustees.
	
	We also observe that the critical value $x_{T,\star}^{(TU)}$ in the well-mixed population is determined only by $R_T$, but in the structured populations it is also related to $R_U$. According to the expressions representing different scenarios outlined in Eq.~(\ref{eq_tudiscuss}), the advantage of trustworthiness instead undermines maintaining a high proportion of trustworthy trustees (i.e., an increase in $R_T$ results in a decrease in $x_{T,\star}^{(TU)}$), while the advantage of untrustworthy trustees facilitates maintaining a high proportion of trustworthy trustees (i.e., an increase in $R_U$ results in an increase in $x_{T,\star}^{(TU)}$). The mechanism behind these counterintuitive phenomena is that, during the evolution of the three strategies, an increase in $R_T$ favors the reproduction of investors, which in turn contributes to the exploitation by untrustworthy trustees, and thus a larger proportion of untrustworthy trustees relative to trustworthy trustees at the extinction of investors. The effect of $R_U$ works vice versa.
	
	Figure~\ref{fig2}(b) numerically compares the critical points of the well-mixed population and the structured population with different updating rules on the equilibrium $TU$-edge, which demonstrates our conclusions intuitively. 
	
	\section{Conclusion}
	Trust evolves between investors, trustworthy and untrustworthy trustees. In the two-player game framework, this work represented the trust game as a $3\times3$ payoff matrix. Under weak selection and pair approximation, we investigated the evolutionary dynamics in a structured population. Our results hold on any regular network structure where each player has the same number of neighbors, including but not limited to lattice graphs and random regular graphs. Analytical solutions under four strategy updating rules, including PC, BD, IM, and DB, were obtained. We found that structured populations do not always favor the evolution of trust.
	
	On the one hand, investors and trustworthy trustees are interdependent to generate a payoff. In this sense, the well-mixed population is most conducive to the coexistence of investors and trustworthy trustees and can maximize the mean payoff of the population. In contrast, the DB, IM, and PC/BD updating rules in structured populations sequentially reduce investors and shift the equilibrium point away from the optimal ratio in coexistence. The principle underlying spatial reciprocity is increasing exposure to the same strategy at the cost of decreasing the chance of meeting with a different strategy, which is unfortunately counterproductive in the context of investor-trustee interplay that requires encountering the opposing strategy to generate a payoff.
	
	On the other hand, the invasion of untrustworthy trustees destroys the collaborative relationship between investors and trustworthy trustees, leading to the extinction of investors. In both well-mixed and structured populations, the system eventually stabilizes in an equilibrium line, where trustworthy and untrustworthy trustees coexist. We find that the PC and BD updating rules in the structured population are most conducive to increasing the stability interval for retaining more trustworthy trustees. Furthermore, the IM and DB updating rules are unfavorable for retaining trustworthiness, respectively, while the well-mixed population is the least favorable.
	
	To sum up, this study supplemented an analytical perspective on the evolution of trust in structured populations. Based on this approach, extensive factors such as punishment, reward, and reputation can be incorporated to modify the payoff matrix and the role of new factors in structured populations can be investigated. Moreover, there have been many different explorations~\cite{kumar2020evolution,sun2022evolution,chiong2022evolution} from those in which investors and trustees can transform into each other, and their versions in structured populations can also be further investigated.
	
	\section*{Declaration of competing interest}
	None.
	
	\appendix
	\renewcommand\thefigure{\Alph{section}\arabic{figure}} 
	\renewcommand{\theequation}{\thesection.\arabic{equation}}
	\section{Stability analysis}\label{sec_appen}
	\setcounter{figure}{0}
	\setcounter{equation}{0}
	Under the constraint $\sum_i x_i=1$, we can substitute for $x_I=1-x_U-x_T$. The system~(\ref{eq_replicator_wm}), (\ref{eq_replicator_pc}), (\ref{eq_replicator_im}), and (\ref{eq_replicator_db}) can be expressed as
	\begin{equation}\label{eq_xtxu}
		\left\{
		\begin{array}{@{\hspace{0.1em}}l@{\hspace{0.25em}}l}
			\dot{x}_T &= \displaystyle{(1-x_T-x_U) x_T t_V \left[R_T+\Delta-x_T (2R_T-1)-x_U (R_U-1)\right]}, \\[1em]
			\dot{x}_U &= \displaystyle{(1-x_T-x_U) x_U t_V \left[R_U+(R_U+1)\Delta-x_T (2R_T-1)-x_U (R_U-1)\right]},
		\end{array}
		\right.
	\end{equation}
	where
	\begin{equation}\label{eq_delta}
		\Delta=\begin{cases}
			\displaystyle{0}, &\mbox{WM,} \\[1em]
			\displaystyle{\frac{1}{k-2}}, &\mbox{PC/BD,} \\[1em]
			\displaystyle{\frac{3}{(k+3)(k-2)}}, &\mbox{IM,} \\[1em]
			\displaystyle{\frac{1}{(k+1)(k-2)}}, &\mbox{DB.}
		\end{cases}
	\end{equation}
	Through this approach, we study the properties of the four systems together, where $\Delta\geq 0$.
	
	To analyze the stability of $\mathbf{x}^{(I)}$ and $\mathbf{x}^{(IT)}$, we compute the Jacobian matrix of the system~(\ref{eq_xtxu}),
	\begin{equation}
		J=\begin{pmatrix}
			\displaystyle{\frac{\partial \dot{x}_T}{\partial x_T}} & 
			\displaystyle{\frac{\partial \dot{x}_T}{\partial x_U}} \\[1em]
			\displaystyle{\frac{\partial \dot{x}_U}{\partial x_T}} & 
			\displaystyle{\frac{\partial \dot{x}_U}{\partial x_U}}
		\end{pmatrix},
	\end{equation}
	where
	\begin{align}
		\frac{\partial \dot{x}_T}{\partial x_T}
		=&~(1-2x_T-x_U)\left[R_T+\Delta-x_U (R_U-1)
		\right]t_V-x_T(2-3x_T-2x_U)(2R_T-1)t_V
		, \\
		\frac{\partial \dot{x}_T}{\partial x_U}
		=&-x_T t_V [(1-x_T)(R_U-1)+R_T+\Delta-x_T(2R_T-1)-2x_U(R_U-1)], \\
		\frac{\partial \dot{x}_U}{\partial x_T}
		=&-x_U t_V [(1-x_U)(2R_T-1)+R_U+(R_U+1)\Delta-2x_T(2R_T-1)-x_U(R_U-1)], \\
		\frac{\partial \dot{x}_U}{\partial x_U}
		=&~(1-x_T-2x_U)\left[R_U+(R_U+1)\Delta-x_T (2R_T-1)
		\right]t_V-x_U(2-2x_T-3x_U)(R_U-1)t_V.
	\end{align}
	
	An equilibrium point is stable if the Jacobian matrix at that point is negative definite. According to basic linear algebra, a matrix is negative definite if all of the odd-ordered principal minors are less than zero and all even-ordered principal minors are greater than zero. This condition can be expressed as: $\partial \dot{x}_T/\partial x_T<0$ and $(\partial \dot{x}_T/\partial x_T)(\partial \dot{x}_U/\partial x_U)-(\partial \dot{x}_T/\partial x_U)(\partial \dot{x}_U/\partial x_T)>0$.
	
	For $\mathbf{x}^{(I)}=(1,0,0)$, we have
	\begin{equation}
		\left.\frac{\partial \dot{x}_T}{\partial x_T}
		\right|_{\mathbf{x}=\mathbf{x}^{(I)}}
		=(R_T+\Delta)t_V>0.
	\end{equation}
	Therefore, $\mathbf{x}^{(I)}$ is unstable.
	
	According to Eqs.~(\ref{eq_xit_wm}) and (\ref{eq_xit_st}), $\mathbf{x}^{(IT)}$ can be expressed as 
	\begin{equation}
		\mathbf{x}^{(IT)}=\frac{1}{2R_T-1}
		\left(R_T-1-\Delta,R_T+\Delta,0\right).
	\end{equation}
	Therefore, for $\mathbf{x}^{(IT)}$, we have
	\begin{equation}\label{eq_Jxit_xtxt}
		\left.\frac{\partial \dot{x}_T}{\partial x_T}
		\right|_{\mathbf{x}=\mathbf{x}^{(IT)}}
		=-(1-x_T^{(IT)})(R_T+\Delta)t_V<0,
	\end{equation}
	which is the reason why $\mathbf{x}^{(IT)}$ is stable along the $IT$-edge: let $x_U=0$, then the reduced 2-strategy system of $I$ and $T$ can be described by $\dot{x}_T$ and the stability can be judged by only Eq.~(\ref{eq_Jxit_xtxt}).
	
	We may compute the even-ordered principal minor to further complete the stability analysis of $\mathbf{x}^{(IT)}$ in the 3-strategy system. However, a quick way to prove that $\mathbf{x}^{(IT)}$ is unstable is by showing $\left.\partial \dot{x}_U/\partial x_U \right|_{\mathbf{x}=\mathbf{x}^{(IT)}}>0$ directly. This is because switching the two equations in the system~(\ref{eq_xtxu}) does not influence its properties: $(\dot{x}_T,\dot{x}_U)$ is equivalent to $(\dot{x}_U,\dot{x}_T)$, whose stability should be ensured by $\partial \dot{x}_U/\partial x_U<0$, the odd-ordered principal minor less than zero. However, we have
	\begin{equation}
		\left.\frac{\partial \dot{x}_U}{\partial x_U}
		\right|_{\mathbf{x}=\mathbf{x}^{(IT)}}
		=(1-x_T^{(IT)})(R_U-R_T+R_U \Delta)t_V>0,
	\end{equation}
	which proves that $\mathbf{x}^{(IT)}$ cannot be stable in the 3-strategy system. To sum up, $\mathbf{x}^{(IT)}$ is a saddle point only stable along the $IT$-edge.
	
	To study the stability of the equilibrium line,
	\begin{equation}
		\mathbf{x}^{(TU)}=\left(
		0, x_T^{(TU)}, x_U^{(TU)}
		\right),
	\end{equation}
	we need to comprehend the physical insight into what happens on the $TU$-edge. Without the investment of $I$-players, the $T$- and $U$-players become indistinguishable in terms of payoff: $f_T=f_U$ when $x_I=0$ according to Eq.~(\ref{eq_f}). Therefore, the system does not evolve on the $TU$-edge, which is the reason why the $TU$-edge is equilibrated everywhere.
	
	In this way, we treat $T$ and $U$ as a whole and reduce the 3-variable system to a 2-variable system: $x_I$ and $x_T+x_U$. According to $\sum_{i}x_i=1$, we can further cancel $x_T+x_U=1-x_I$, $x_U=1-x_I-x_T$, and express the replicator dynamics by only $\dot{x}_I$ with an input parameter $x_T$,
	\begin{align}\label{eq_xi}
		\dot{x}_I
		=&~x_I t_V \Big\{
		x_T (R_T-1)-(1-x_I-x_T)-[x_T+(1-x_I-x_T)(R_U+1)]\Delta \nonumber\\
		&-x_I x_T (2R_T-1)-x_I (1-x_I-x_T)(R_U-1)
		\Big\},
	\end{align}
	where $\Delta$ has the same meaning in Eq.~(\ref{eq_delta}).
	
	The single-order Jacobian matrix of the system~(\ref{eq_xi}) is
	\begin{align}
		\frac{\mathrm{d} \dot{x}_I}{\mathrm{d} x_I}
		=&~t_V \Big\{x_T (R_T-1)-(1-2x_I-x_T)-[x_T+(1-2x_I-x_T)(R_U+1)]\Delta \nonumber\\
		&-2x_I x_T (2R_T-1)-x_I(2-3x_I-2x_T)(R_U-1)\Big\}.
	\end{align}
	To analyze the stability of $\mathbf{x}^{(TU)}=\left(0, x_T^{(TU)}, x_U^{(TU)}\right)$, we compute
	\begin{equation}
		\left.\frac{\mathrm{d} \dot{x}_I}{\mathrm{d} x_I}
		\right|_{\mathbf{x}=\mathbf{x}^{(TU)}}
		=t_V \Big\{x_T^{(TU)} (R_T-1)-(1-x_T^{(TU)})-[x_T^{(TU)}+(1-x_T^{(TU)})(R_U+1)]\Delta\Big\},
	\end{equation}
	from which we know that $\mathbf{x}^{(TU)}$ is stable if 
	\begin{equation}\label{eq_xtucondi}
		\left.\frac{\mathrm{d} \dot{x}_I}{\mathrm{d} x_I}
		\right|_{\mathbf{x}=\mathbf{x}^{(TU)}}<0
		\Leftrightarrow
		x_T^{(TU)}<\frac{1+(R_U+1)\Delta}{R_T+R_U \Delta}
		\equiv x_{T,\star}^{(TU)}.
	\end{equation}
	
	Eq.~(\ref{eq_xtucondi}) gives the explicit expression of $x_{T,\star}^{(TU)}$ that we presented in Eqs.~(\ref{eq_xtu_wm}) and (\ref{eq_xtu_st}). On the equilibrium line $\mathbf{x}^{(TU)}$, the interval of $x_T^{(TU)}<x_{T,\star}^{(TU)}$ is stable.
	
	Finally, we note that an equilibrium point on the $IU$-edge can be obtained, but it is not valid. Solving the system~(\ref{eq_xtxu}), we find the following solution:
	\begin{equation}
		\mathbf{x}^{(IU)}=\frac{1}{R_U-1}
		\left(-1-(R_U+1)\Delta,0,R_U+(R_U+1)\Delta \right).
	\end{equation}
	However, since $x_I^{(IU)}=[-1-(R_U+1)\Delta]/(R_U-1)<0$, it does not exist, which is why it was omitted from the main text.


\end{document}